\documentclass[useAMS,usenatbib]{mn2e}

\usepackage{epsfig,amsmath,amssymb}
\title[Extended Magnetosphere]{The Extended Pulsar Magnetosphere}
\author[Kalapotharakos et al.]{Constantinos Kalapotharakos$^{1,2}$\thanks{E-mail:
constantinos.kalapotharakos@nasa.gov (CK); icontop@academyofathens.gr (IC); demos.kazanas@nasa.gov (DK)} and Ioannis
Contopoulos$^{3}$ and Demos Kazanas$^{2}$\\
$^{1}$University of Maryland, College Park (UMDCP/CRESST), College Park, MD 20742, USA\\
$^{2}$Astrophysics Science Division, NASA/Goddard Space Flight Center, Greenbelt, MD 20771, USA\\
$^{3}$Research Center for Astronomy and Applied Mathematics, Academy of Athens, Athens 11527, Greece}
\begin{document}

\date{Accepted ??. Received 2011 September 22; in original form 2011 June 3}

\pagerange{\pageref{firstpage}--\pageref{lastpage}} \pubyear{2011}

\maketitle

\label{firstpage}

\begin{abstract}
We present the structure of the 3D ideal MHD pulsar magnetosphere to
a radius ten times that of the light cylinder, a distance about an
order of magnitude larger than any previous such numerical
treatment. Its overall structure exhibits a stable, smooth,
well-defined undulating current sheet which approaches the kinematic
split monopole solution of Bogovalov~1999 only after a careful
introduction of diffusivity even in the highest resolution
simulations. It also exhibits an intriguing spiral region at the
crossing of two zero charge surfaces on the current sheet, which
shows a destabilizing behavior more prominent in higher resolution
simulations. We discuss the possibility that this region is
physically (and not numerically) unstable. Finally, we present the
spiral pulsar antenna radiation pattern.
\end{abstract}


\section{Introduction}

Almost half a century since the discovery of pulsars and despite the
great observational progress, a comprehensive understanding of the
observed radiation remains elusive. This is to a large extent due to
its breadth in frequency (from radio to $\gamma-$rays) and
dependence on the pulsar rotation phase as well as on its potential
dependence on the (unknown) observer inclination angle and the angle
between the rotation and magnetic axes.

Pulsar radio emission remains the hardest to model (mainly because
of its apparently coherent character) and most models that claim to
account for its main features (multiple pulse profiles, polarization
angle sweeps, pulse width-to-wavelength relation, etc.) are simply
phenomenological. It is considered that radio emission originates in
(probably hollow) conical beams above the magnetic polar caps. The
situation is slightly better at higher energies (optical to
gamma-rays) where the recent Fermi observations indicate the
emission region to be associated with the outer magnetosphere even
beyond the light cylinder \citep{Ransom10,Guill11}. The exact
emission mechanism remains unclear, however, it is generally
considered to be curvature radiation emitted by relativistic
electrons moving along the pulsar magnetic field lines near the
separatrix between the closed and open field lines, although
synchrotron and Inverse Compton emission could potentially make
important contributions to the spectrum \citep{MH04,AKH08}.

Given the relativistic unidirectional motion of the pulsar radiating
particles, any significant advance in understanding their emission
across the electromagnetic spectrum must take into account the
detailed structure of the pulsar magnetosphere. Its structure, even
in the axisymmetric, aligned dipole field had been unknown until
rather recently. Following the original solution of Contopoulos,
Kazanas \& Fendt~(1999), a number of additional treatments have
established the precise structure of this configuration
(Gruzinov~2005; Contopoulos~2005; Timokhin~2006). More recently the
structure of the non-aligned dipole was obtained numerically
(Spitkovsky~2006; Kalapotharakos \& Contopoulos~2009); this was then
used to produce kinematic model light curves by placing the emission
of radiation both near to (Bai \& Spitkovsky~2010) and well outside
the light cylinder \citep{CK10}, with results that apparently are
both consistent with observations. The idea that the observed
radiation is sited outside the light cylinder was not new. Lyubarsky
(1996) had proposed the possibility that the high energy emission
comes from sites just beyond the light cylinder at the reconnection
region near the Y-point while Petri \& Kirk (2005) proposed the
stripped wind model in which the observed radiation comes from a
zone far beyond the light cylinder.

The most recent multi-frequency observations of pulsar emission do
not seem to provide a preference between these rather diverse sites
of pulsar emission, being in general phenomenological agreement with
both. In fact, in most pulsars with both gamma-ray and radio
observations available (56 as of April 2011), the radio pulse
precedes the main gamma-ray pulsar by about 15\% to 40\% of the
period \citep{AbdoPCat}. This is consistent with the radio emission
being associated with the polar cap, and the high energy emission
being associated with a region further out along the open field
lines. Still, 7 pulsars (the Crab and another 6 out of 20 known
millisecond gamma-ray pulsars) do not follow this simple picture. In
these, the radio and high energy peaks are in phase. This suggests
that, at least in these 7 pulsars, radio emission is associated with
the extended outer magnetosphere. Interestingly enough, in the case
of the Crab pulsar, a radio component associated with the polar cap
emission region has also been observed, albeit only at low
frequencies (Moffett \& Hankins~1996). This may further suggest
that, at least under certain circumstances, radio emission may be
coming both from the polar cap and the outer magnetosphere.

This issue as well as the general stability of the magnetosphere at
larger distances motivate a study of its structure at larger radii.
Furthermore, a theoretical model that purports to account for the
origin of radio emission in the outer magnetosphere was first
proposed by Ardavan~(1998) (see also Ardavan {\em et al.}~2008).
Ardavan proposed that a superluminally corotating pattern of
electric charges and currents generates at large distances regions
of enhanced electromagnetic radiation in vacuum where the local
Poynting flux drops as the inverse ({\em and not} the inverse
square) of the distance from the source. We were particularly
intrigued by this proposition and decided to investigate its
implications in the pulsar magnetosphere through extensive numerical
simulations that probe its structure well outside the light
cylinder. As we will see, our results, although still not 100\%
conclusive, do not seem to support the Ardavan model.

The study of the outer magnetosphere (up to 10$r_{lc}$; $r_{lc}$ is
the light cylinder radius) requires, on one hand high enough
resolution simulations in order to guarantee minimal numerical
dissipation, and on the other the introduction of some (small)
amount of diffusivity in order to reduce noise and stabilize the
numerical scheme. These simulations exhibit a stable and
well-defined undulating smooth current sheet tending towards the
split monopole solution (Bogovalov~1999). However, the non-diffusive
simulations exhibit a destabilizing spiral region at the crossing of
two zero charge (null) surfaces that happen also to be located on
the current sheet. This destabilizing behavior is enhanced in
simulations of increased resolution suggesting a physical (and not
numerical) instability.

This paper is structured as follows: in \S~2 we discuss our
numerical procedure in comparison with previous treatments of the
problem. In \S~3 we present the pulsar antenna radiation pattern.
Finally, our conclusions are summarized in \S~4.

\section{Numerical simulations}

\subsection{Procedure}

The 3D force-free structure of the pulsar magnetosphere in ideal MHD
has been investigated numerically only very recently by two groups
(Contopoulos \& Kalapotharakos 2010, hereafter CK10, and Bai \&
Spitkovsky 2010). Both investigations have been based on the
formalism of Force-Free Electrodynamics (hereafter FFE;
Gruzinov~1999; Blandford~2002) and on the numerical procedure
described by Spitkovsky~(2006). The two investigations differ in the
outer boundary conditions of the computational grid. CK10
implemented a Perfectly Matched Layer (PML) that guarantees
absorption/non-reflection of the waves reaching the edge of the
computational domain. This allowed them to run their original 3D
pulsar magnetosphere simulation in a small computational box
extending out to only about 2.5 light cylinder radii from the `star'
for as long as necessary to achieve steady state. On the other hand,
Spitkovsky~(2006) and Bai \& Spitkovsky~(2010) did not implement any
special outer boundary condition but simply used a much larger
computational domain; as such, they followed the evolution of the
pulsar magnetosphere in the inner 2 light cylinder radii, for only
about 2 stellar rotations (that is until the waves reflected at the
boundary reach the distance of 2 light cylinder radii). Both
investigations reached a very similar steady state magnetospheric
configuration.

The issues raised in the Introduction suggest the need for computing
the structure of the magnetosphere on a more extended domain than
employed so far. Obviously, the implementation of a PML makes our
code more suitable to address this problem with the same amount of
computational resources than any other FFE numerical code currently
available (see CK10 for details about the code). We have run
simulations extending out to 10 light cylinder radii in all
directions for various pulsar inclination angles. Each simulation
required about 35 hours on a 1000-core supercomputer. We start off
with an inclined magnetostatic dipole field at the center of our
cartesian computational cube. At time $t=0$, we set the central
dipole in steady rotation. We run the simulation for about 2 stellar
rotations, which is the time needed for the initial transient wave
to reach the outer boundary of the numerical grid plus another 2
rotations to safely reach steady state across our computational
grid. The resolution of the simulation is 50 grid points per light
cylinder radius, i.e. two times higher than that in CK10 (for a
total of $1000^3$ grid cells - not taking into account those of the
PML). We would like to note here that, in distinction to our
original smaller scale simulations that reached only out to
2.5~$r_{lc}$, our present extended ones include artificial
diffusion. This was required in order to better approach the final
steady state(we will return to this important point in \S~2.3 and in
the end of \S~3). More specifically we added artificial numerical
diffusion terms in the time-dependent Maxwell equations
\begin{eqnarray}
\label{maxwelldif}
  \frac{1}{c}\frac{\partial \mathbf{B}}{\partial t} &=&
  -\nabla \times \mathbf{E}
  +\eta\  r_{lc}\nabla^2\mathbf{B} \\
  \frac{1}{c}\frac{\partial \mathbf{E}}{\partial t} &=&
  \nabla \times \mathbf{B}
  -\frac{4\pi}{c}\mathbf{J}+\eta\ r_{lc}\nabla^2\mathbf{E}
\end{eqnarray}
where $\eta$ is a dimensionless diffusion parameter acting mostly in
the outer magnetosphere beyond 1.5~$r_{lc}$. The value of $\eta$ is
weighted with $\left|\nabla^2\mathbf{B}\right|$ and
$\left|\nabla^2\mathbf{E}\right|$, and its minimum and maximum
values throughout the numerical grid are $2\times 10^{-4}$ and
$6\times 10^{-3}$ respectively.

\begin{figure*}
\includegraphics[width=13.7cm]{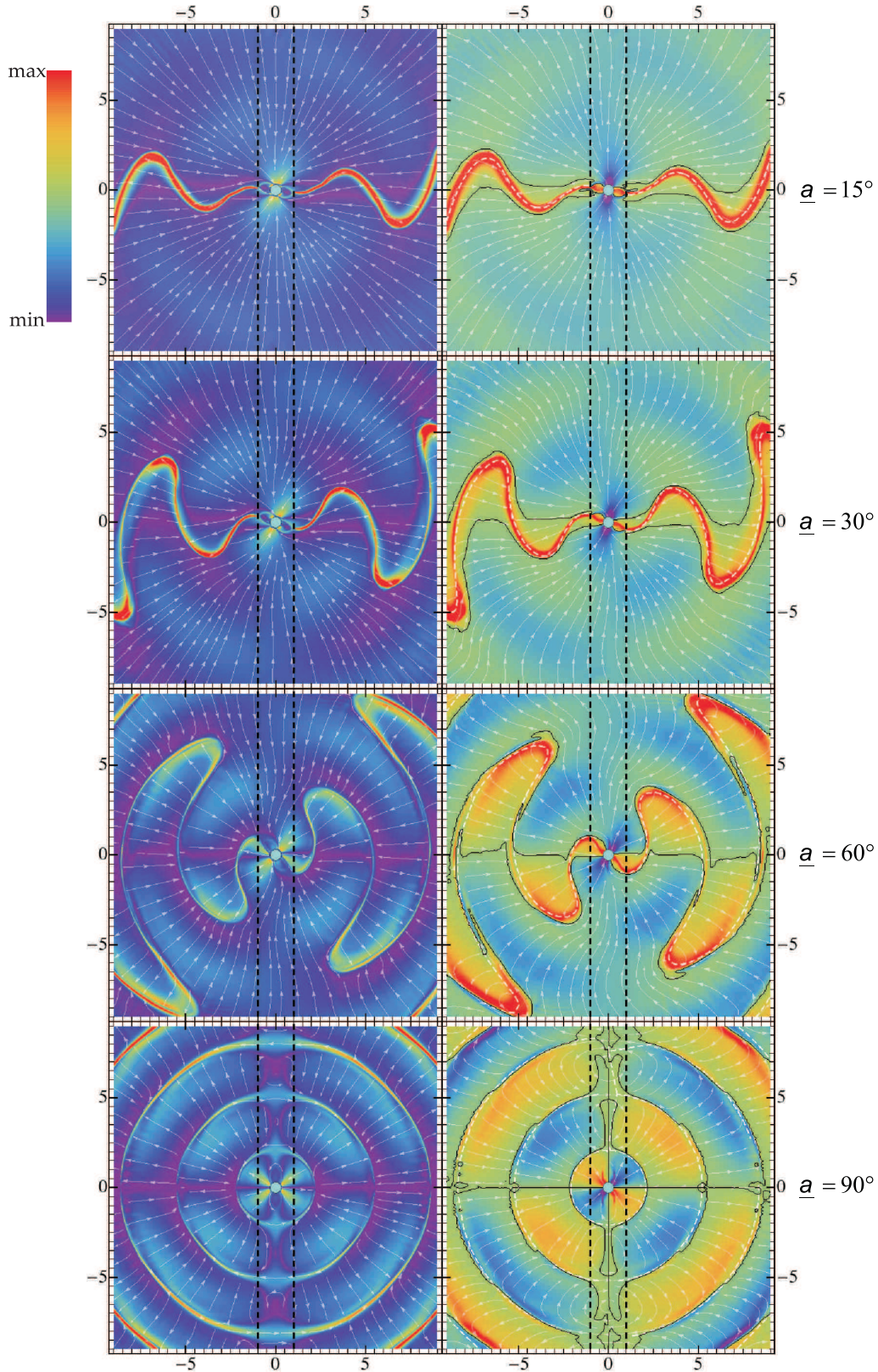}
\caption{Meridional cross section of the magnetosphere on the
poloidal ($\bf \mu,\Omega$) plane in steady state for various pulsar
inclination angles. Left panels: meridional electric current density
multiplied by $r^2$ (color plot) and direction (arrows). Right
panels: electric charge density multiplied by $r^2$ (color plot) and
meridional magnetic field direction (arrows). Inclination angles
from top to bottom:$15^\circ$, $30^\circ$, $60^\circ$, $90^\circ$.
We have considered the mean values of all the plotted quantities
over one period in order to reduce the noise level. Light cylinder
shown with black vertical dashed lines. The black solid lines and
the white dashed lines in the right hand column indicate the zero
charge lines and the corresponding section of the original
`Bogovalov' type current sheet, respectively.}
\end{figure*}

\begin{figure*}
\includegraphics[width=\textwidth]{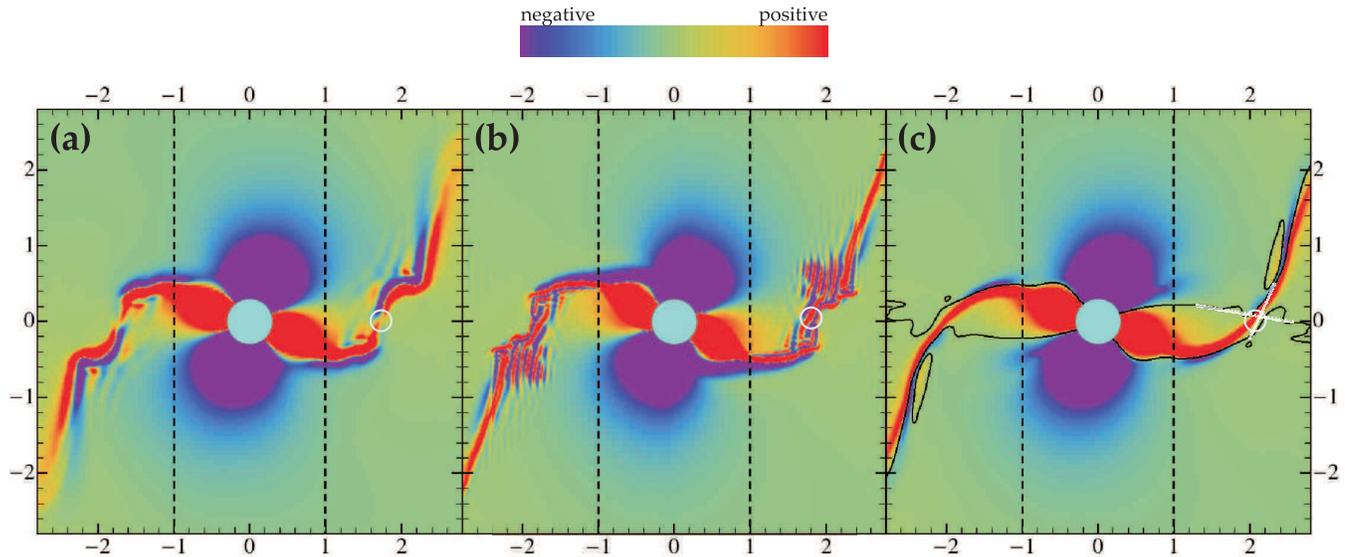}
\caption{Detail of the electric charge density in the poloidal ($\bf
\mu,\Omega$) plane in steady state for an inclination angle of
$30^\circ$. Left panel: low resolution (25 grid points per light
cylinder radius) non-diffusive simulation. Center panel: high
resolution (50 grid points per light cylinder radius) non-diffusive
simulation. Right panel: high resolution (50 grid points per light
cylinder radius) simulation with artificial diffusion as in Eqs.~(1)
\& (2). Zero charge (null) surfaces shown with solid black lines.
Crossing of null surfaces shown with white line elements and circle.
Light cylinder shown with dashed lines.}
\end{figure*}

\subsection{Results}

Our results are summarized in Fig.~1 where we plot various physical
quantities on the poloidal ($\bf\mu,\Omega$) plane (i.e. the plane
that contains both the rotation and magnetic axes) in steady state
for various pulsar inclination angles. These plots were obtained
after about 3-4 stellar rotations, beyond which an almost steady
pattern of currents and fields that corotates with the central star
is established. Note that in Fig.~1 we have considered the mean
values of all the plotted quantities over one period in order to
reduce significantly the noise level. On the left, we plot the
distribution of meridional electric current density multiplied by
$r^2$ (color plot) and direction (arrows), and on the right the
electric charge density multiplied also by $r^2$ (color plot) and
meridional magnetic field direction (arrows). The black lines in the
right hand column denote the zero charge lines. The inclination
angles are $15^\circ$, $30^\circ$, $60^\circ$, and $90^\circ$ from
top to bottom respectively. Note that, in FFE, when a corotating
steady-state is reached, the poloidal electric current and magnetic
field components should be parallel everywhere. We do acknowledge
that this condition is satisfied to a high degree in our simulations
but obviously not 100\%. The situation was much worse in our lower
resolution -25 grid points per light cylinder radius- simulations.
We offer some idea about the cause of this effect in the end of
\S~3.

The most noticeable element of our solution is that an undulating
equatorial current sheet forms as predicted by Bogovalov~(1999,
hereafter B99). This is the 3D generalization of the equatorial
current sheet discovered in the axisymmetric investigation of
Contopoulos, Kazanas \& Fendt~(1999). The white dashed line in the
right hand column of Fig.~1 indicate the section of the original
`Bogovalov' type current sheet on the ($\bf\mu,\Omega$) plane. The
current sheet survives as far as the outer boundaries of our
numerical simulation for several stellar rotations and is not
destroyed by instabilities or reconnection (see however also \S~2.3,
3). It originates at the positions where the closed line region
`touches' the light cylinder. The current density at those positions
is clearly non-uniform. The opening half-angle of the current sheet
above and below the rotational equator is about equal to the pulsar
inclination angle, as predicted by B99.

\subsection{Numerical or physical instabilities?}

CK10 and \citet{BS10} have already identified regions of enhanced
electric charge density double layers (positive and negative)
described as `knots' just outside the light cylinder. Obviously, the
electric field is greatly enhanced between such layers. The results
presented in CK10 did not include any artificial diffusivity. In the
present study we realized that this behavior is more prominent in
higher resolution non-diffusive simulations. In Fig.~2 we plot
details of the electric charge density distribution in the poloidal
plane $(\bf \mu,\Omega)$ corresponding to $a=30^\circ$ as obtained
in the low (Fig.~2a) and high (Fig.~2b) resolution non-diffusive
simulations, and in the high resolution simulation with artificial
numerical diffusivity (Fig.~2c). In Fig.~2a we observe that the
region of strong positive and negative electric charges is smoother
than in Fig.~2b. As a general trend, in higher resolutions numerical
instabilities tend to be suppressed, whereas physical instabilities
tend to manifest themselves. The fact that the feature observed in
Fig.~2a is enhanced in simulations of increased resolution signifies
an instability of physical rather than numerical underpinnings.
Nevertheless, we cannot yet rule out that it may also be due to some
kind of numerical instability. In Fig.~2c we see that the
introduction of the diffusion term in eqs.~(1), (2) clearly smooths
out this feature.

We realized that such `knots' appear where two different zero charge
surfaces cross each other {\em on} the current sheet. In Fig.~2 the
sign of the charge density on the poloidal plane reveals this
strange topology: the crossing of two zero charge (null) surfaces
creates an X-point on the current sheet. Above and below this point
along the current sheet double layers of opposite electric charges
are formed. We have also found that dissipative solutions of the
magnetosphere above and below this point exhibit electric field
components parallel and antiparallel to the magnetic field,
respectively.

\begin{figure*}
\includegraphics[width=\textwidth]{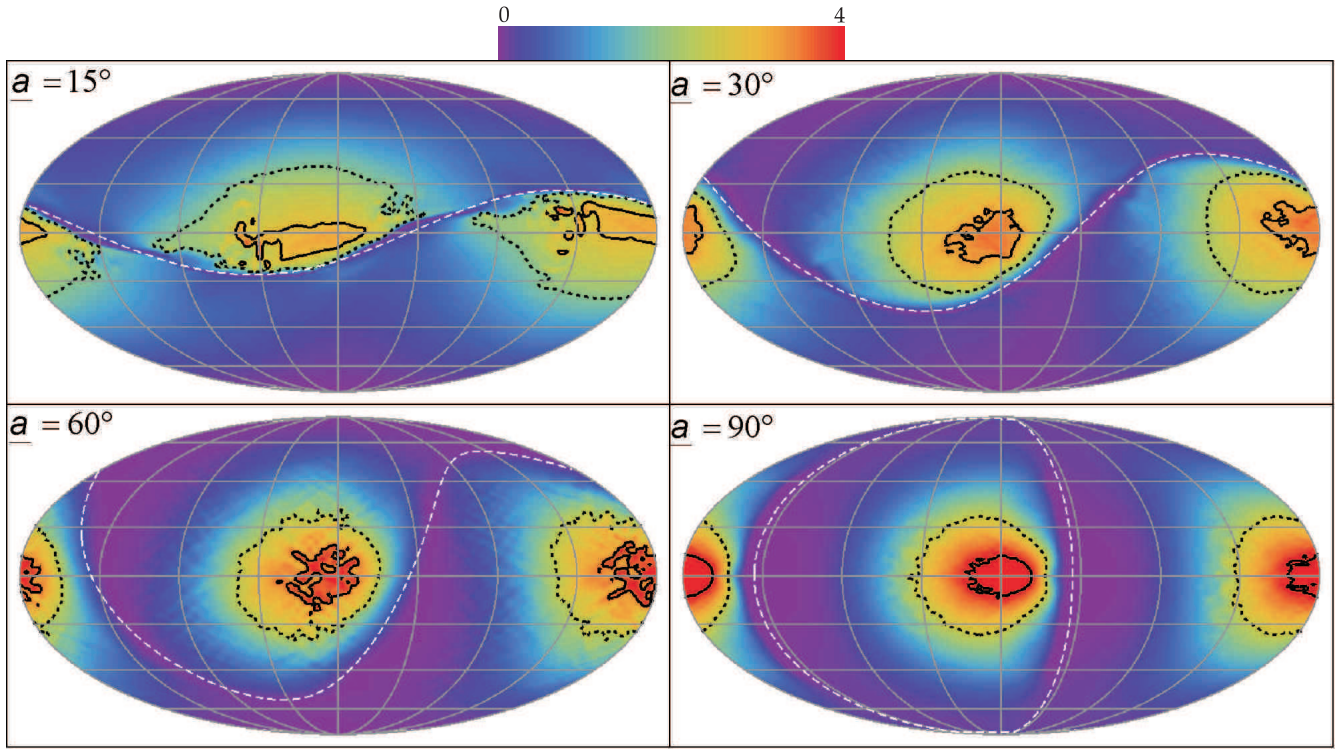}
\caption{Distribution of the Poynting flux in  Mollweide
(elliptical) projection of the sphere $r=8 r_{lc}$ centered on the
central star for the inclination angles indicated in the panels. The
linear color scale ranges from 0 up to 4 times the corresponding
mean Poynting flux value. The highest $50\%$ of the values are
contained inside the dashed black lines, and the highest $10\%$
inside the solid black lines. The maximum of the Poynting flux forms
a spiral pattern near the (rotational) equator. Note that the 50\%
of the Poynting flux in $a=90^{\circ}$ case (dotted curve) is
emitted only through 15\% of the entire $4\pi$ solid angle. The same
area corresponding to the $a=15^{\circ}$ case is 40\% larger. The
white dashed lines trace the locus of the current sheet.}
\end{figure*}

The question with this feature is whether it represents a real
structure or an artifact of our numerical scheme. Is this feature
introduced by the  non-diffusive scheme (Fig.~2a,b) or is it that
the diffusion introduced in producing Fig.~2c smooths out a really
singular feature of the solution? The fact that two independent
numerical schemes (CK10; Bai \& Spitkovsky 2010) exhibit similar
behavior and that this is a rather extraordinary region (the
confluence of 2 zero charge surfaces on a current sheet) supports
the view that this may indeed be a real feature.

\section{The pulsar antenna radiation pattern}

Overall, the magnetic field assumes an open split-monopole-like
structure that corresponds to the radiation field of an antenna with
${\bf E}$ almost perpendicular to ${\bf B}$, and both $E$ and $B$
dropping like $1/r$ on average
with distance from the central antenna (in our case, the rotating
central dipole). We remind the reader that in any radiation pattern,
a characteristic distance exists that separates the inner from the
outer radiation field region, the so-called Fresnel distance defined
as
\begin{equation}
r_{\rm Fresnel}\equiv l^2/\lambda\sim r_{lc}\ .
\end{equation}
Here, $r_{lc}\equiv cP/(2\pi)$ is the light cylinder radius;
$l\sim 2r_{lc}$ is the linear extent of the `pulsar antenna'; and
$\lambda= 2\pi r_{lc}$ is the wavelength of the radiation as seen
in the electromagnetic field patterns of Fig.~1. Our simulation
box extends beyond a few times the above distance, and therefore,
it is suitable to study the pulsar antenna far-field radiation
pattern.

While the Poynting flux of the rotating magnetosphere averaged over
the entire $4 \pi$ solid angle decreases as $\sim 1/r^2$, as
expected from the conservation of electromagnetic radiation, this
flux is not distributed uniformly over the sphere. In Fig.~3 we plot
the Poynting flux distribution on the Mollweide (elliptical)
projection of a sphere of radius $r=8 r_{lc}$ centered on the
central star ($\theta=0$ along the axis of rotation). One sees that
the Poynting flux displays local maxima. Obviously, in steady-state,
the radiation pattern should corotate with the star. For
$a=90^\circ$ the 50\% of the Poynting flux (black dotted curve of
Fig.~3) is emitted from only $0.6\pi$~sr.

Ardavan (1998) proposed that singular electromagnetic radiation
features take place in vacuum when a pattern of electric charges
and currents corotates superluminally (only the pattern corotates;
nothing is moving superluminally). In the Introduction, we
acknowledged that this was one of our original motivations for
extending the simulations of CK10 to much larger distances.
Preliminary results hinted that the local Poynting flux in regions
of enhanced electromagnetic field discussed in \S~2.3 drops as
$1/r$ (and {\em not} as $1/r^2$) with distance $r$, as anticipated
in vacuum by Ardavan~(1998). This was very encouraging! However,
the fact that certain features of the simulation were noisy and
not consistent with steady-state led us to increase the spatial
resolution by a factor of two and to introduce diffusivity in
order to stabilize the numerical scheme. As a consequence any
singular electromagnetic radiation feature was killed.

Unfortunately, we are still not in a position to completely
dismiss the possibility that the physical solution includes
regions of enhanced electromagnetic field (radiation caustics?)
and that the introduction of (albeit small) diffusion does not
artificially suppress their development. Furthermore, diffusivity
introduces also a gradual decrease of $E$ with distance below its
theoretical steady-state corotation value $(r\sin\theta
/r_{lc})B_p$, and therefore, magnetic field lines gradually loose
corotation with the central star. This manifests itself in Fig.~1
through the inability of our numerical procedure to satisfy 100\%
the steady-state requirement that the poloidal components of the
electric current and the magnetic field must be parallel
everywhere.

\section{Conclusions}

In this paper, we presented the results of extensive numerical
simulations of the 3D force-free ideal pulsar magnetosphere out to
a distance of about 10 times the light cylinder radius. Our
results clearly show that the undulating equatorial current sheet
survives over that distance and over several stellar rotations,
implying that it represents an almost steady feature of the
magnetosphere. The introduction of a small artificial diffusivity
was necessary for the stabilization of the numerical scheme. Using
this scheme the current sheet appears smooth and very similar to
the one given analytically in B99. Overall, it is primarily
charged positively (negatively) for an aligned (counter-aligned)
rotator\footnote{Obviously, the concept of alignment gradually
loses its meaning as we approach $90^\circ$ inclination angles.}.

Interestingly enough, the regions of strong charge density first
seen in CK10 are enhanced in higher resolution non-diffusive
simulations. These regions are formed near the locus of the crossing
of two zero-charge surfaces. A destabilizing behavior more prominent
in high resolution simulations is observed around these regions.
Nevertheless, this behavior is suppressed after the introduction of
artificial diffusivity (necessary for the stabilization of the
numerical scheme), thus raising questions about its true nature
(physical or numerical).

We present also the pulsar antenna spiral (for oblique rotators)
radiation pattern. The Poynting flux is not uniformly distributed
over the entire sphere. Its maximum is along a spiral direction
near the (rotational) equator. The Poynting flux corresponding to
the smooth high resolution solutions produced by the diffusive
simulations drops everywhere as $r^{-2}$ implying that enhanced
electromagnetic fields similar to those proposed by Ardavan (1998)
are not present. Their existence, however, cannot yet be
conclusively ruled out since such effects are expected to be
sensitive to diffusivity. Further investigation with much higher
resolution is required.

\section*{Acknowledgments}

We would like to acknowledge extensive discussions with Houshang
Ardavan, as well as with Christos Efthymiopoulos, Avi Loeb, Maxim
Lyutikov, Cole Miller, Ramesh Narayan, and Sasha Tchekhovskoy. We
thank also the anonymous referee whose suggestions motivated us to
develop a novel numerical technique that improved the quality of the
presented results.

\label{lastpage}

\end{document}